\documentclass[a4paper,fleqn,usenatbib]{mnras}

\usepackage[T1]{fontenc}
\usepackage{ae,aecompl}

\usepackage{graphicx}	
\usepackage{amsmath}	
\usepackage{amssymb}	
\usepackage{color}

\definecolor{grey}{rgb}{0.7,0.7,0.7}

\title[The LyC photon production efficiency]{The Lyman-continuum photon production efficiency in the high-redshift Universe}

\author[Stephen M. Wilkins et al.]{
Stephen M. Wilkins,$^{1}$\thanks{E-mail: s.wilkins@sussex.ac.uk}
Yu Feng,$^{2,3}$
Tiziana Di-Matteo,$^{2}$
Rupert Croft,$^{2}$\newauthor
Elizabeth R. Stanway,$^{4}$
Rychard J. Bouwens,$^{5}$
Peter Thomas$^{1}$
\\
$^1$\,Astronomy Centre, Department of Physics and Astronomy, University of Sussex, Brighton, BN1 9QH, UK \\
$^2$\,McWilliams Center for Cosmology, Carnegie Mellon University, Pittsburgh PA, 15213, USA \\
$^3$\,Berkeley Center for Cosmological Physics, University of California, Berkeley, Berkeley CA, 94720, USA \\
$^4$\,Department of Physics, University of Warwick, Gibbet Hill Road, Coventry, CV4 7AL, UK \\ 
$^5$\,Leiden Observatory, Leiden University, NL-2300 RA Leiden, Netherlands \\
}

\date{Accepted XXX. Received YYY; in original form ZZZ}

\pubyear{2015}

\begin{document}
\label{firstpage}
\pagerange{\pageref{firstpage}--\pageref{lastpage}}
\maketitle


\begin{abstract}

The Lyman Continuum photon production efficiency ($\xi_{\rm ion}$) is a critical ingredient for inferring the number of photons available to reionise the intergalactic medium. To estimate the theoretical production efficiency in the high-redshift Universe we couple the BlueTides cosmological hydrodynamical simulation with a range of stellar population synthesis models. We find Lyman Continuum photon production efficiencies of $\log_{10}(\xi_{\rm ion}/{\rm erg^{-1}\, Hz})\approx 25.1-25.5$ depending on the choice of stellar population synthesis model. These results are broadly consistent with recent observational constraints at high-redshift though favour a model incorporating the effects of binary evolution.

\end{abstract}

\section{Introduction}

Ascertaining the sources of the photons responsible for the cosmic reionisation of hydrogen remains a key goal of modern extragalactic astrophysics and a source of continued debate in the literature (e.g. Wilkins et al. 2011a; Robertson et al. 2015; Madau \& Haardt 2015; and Bouwens et al. 2015b). Constraints on the evolution of filling factor of ionised hydrogen ($Q_{\textsc{hii}}$) (see Bouwens et al. 2015b for a recent overview) now suggest that the ionising emissivity evolves similarly to the UV continuum luminosity density hinting at a common source. 

The observed UV luminosity density $\rho_{\textsc{uv}}$ and the ionising emissivity $\dot{n}_{\rm ion}$ of galaxies are connected through the LyC photon production efficiency $\xi_{\rm ion}$ and the escape fraction of Lyman continuum (LyC) photons and UV photons ($f_{\rm esc, LyC}$ and $f_{\rm esc, uv}$ respectively) (Robertson et al. 2013; Kuhlen \& Faucher-Gigu{\`e}re 2012),

\begin{equation}
\dot{n}_{\rm ion} = f_{\rm esc, LyC}\, \xi_{\rm ion}\, \frac{\rho_{\rm uv}}{f_{\rm esc, uv}}.
\end{equation}

The production efficiency $\xi_{\rm ion}$ relates the intrinsic number of LyC photons produced to the UV luminosity. For individual stars these quantities are sensitive to the star's mass, age, chemical composition, rotation (e.g. Topping \& Shull 2015), and whether there are any binary interactions (e.g. Stanway, Eldridge, \& Becker 2015). The production efficiency of a star formation dominated galaxy is then dependent on the joint distribution of these quantities. The production efficiency can thus be predicted using a stellar population synthesis (SPS) model for a given star formation and metal enrichment history and initial mass function (IMF). By including observations of the UV continuum slope $\beta$ (e.g. Wilkins et al. 2011b, Bouwens et al. 2014, Wilkins et al. 2016) it is possible to constrain some of these assumptions (Robertson et al. 2013; Duncan \& Conselice 2015; Bouwens et al. 2015b).

$\xi_{\rm ion}$ can also be constrained observationally using measurements of the nebular line emission combined with gas density and metallicity assumptions. Stark et al. (2015) used measurements of the flux in the C{\sc iv}$\lambda 1548$ line in a lensed Lyman-break galaxy (LBG) at $z\approx 7$ to find $\log_{10}(\xi_{\rm ion}/{\rm erg^{-1}\, Hz})=25.68^{+0.27}_{-0.19}$. More recently Bouwens et al. (2016) used the {\em Spitzer}/IRAC fluxes to constrain the H$_{\alpha}$ emission in a sample of spectroscopically confirmed LBGs at $z=4-5$ (see also Smit et al. 2016). 

In this study we couple six SPS models with the BlueTides hydrodynamical simulation to predict the LyC photon production efficiency. We begin, in Section \ref{sec:BT} by describing the BlueTides simulation. In Section \ref{sec:xi} we investigate the prediction production efficiency as a function of stellar mass (\S\ref{sec:xi.smass}), redshift (\S\ref{sec:xi.z}), and choice of SPS model (\S\ref{sec:xi.SPS}). We then present our conclusions in Section \ref{sec:c}.

\section{The BlueTides Simulation}\label{sec:BT}

BlueTides (see Feng et al. 2015ab for a full description of the simulation) was carried out using the Smoothed Particle Hydrodynamics code {\sc MP-Gadget} with $2\,\times\, 7040^{3}$ particles using the Blue Waters system at the National Centre for Supercomputing Applications. The simulation evolved a $(400/h)^{3}\,{\rm cMpc^3}$ cube from the primordial mass distribution to $z=8$ utilising the {\em Wilkinson Microwave Anisotropy Probe} year 9 cosmological parameters\footnote{($\Omega_{\Lambda}=0.7186$, $\Omega_{\rm matter}=0.2814$, $\Omega_{\rm baryon}=0.0464$, $h=0.697$)} (Hinshaw et al. 2013). BlueTides is the largest (in terms of memory usage) cosmological hydrodynamic simulation carried out. 

Galaxies were selected using a friends-of-friends algorithm at a range of redshifts (though in this study we concentrate on systems at $z<11$). At $z=10/9/8$ there are $14,221/50,713/159,835$ objects with stellar masses with greater than $10^{8}\,{\rm M_{\odot}}$ (i.e. consisting of at least approximately 100 star particles).

The stellar mass function of galaxies in the simulation at $z=8$ closely matches (see Feng et al. 2015ab; Wilkins et al. {\em in-prep} for wider predictions of the simulation) recent observational constraints (e.g. Song et al. 2015). The UV luminosity function (at $z=8-10$) is also consistent with recent observations (e.g. Oesch et al. 2014; Bouwens et al. 2015a; Ishigaki et al. 2015; McLeod et al. 2015; Finkelstein et al. 2015) once a dust attenuation is added to the most luminous systems. Galaxies in the  simulation have a naturally arising rapidly increasing star formation histories the shape of which are largely independent of stellar mass. There is also a strong relationship between the average stellar metallicity and the stellar mass of individual galaxies.

\subsection{Stellar Population Synthesis Modelling}

To estimate the LyC production efficiency we couple the BlueTides simulation with an stellar population synthesis (SPS) model. SPS models combine an evolution model (which gives the temperature and luminosity of each star at a given age and mass) with an atmosphere model (which relates these theoretical values to observable spectral energy distributions). Depending on the choices for each of these components the resulting spectral energy distribution (and thus the LyC production efficiency assuming the same star formation and chemical enrichment history) can vary significantly.

In this work we utilise five SPS models to produce six scenarios (listed in Table \ref{tab:main_tab}). In the case of the {\sc bpass} models, two scenarios were considered. In the first, stars evolved without interaction as in a more traditional SPS code. In the second, the effect of binary interactions on stellar evolution are also  considered, with stars selected from a distribution in initial binary period as well as initial mass, such that the fraction of interacting binaries matches local constraints (see Stanway, Eldridge, \& Becker (2015), Eldridge \& Stanway 2012, Eldridge et al. {\em in-prep}).

In order to provide a direct comparison we assume the same initial mass function (IMF) for each model. For this we choose the Salpeter (1955) IMF over the range $0.1-100\,{\rm M_{\odot}}$. Assuming an alternative IMF is likely to shift the predicted UV luminosity and ionising photon production (and thus, potentially, the production efficiency). However, for changes to the low-mass ($<1\,{\rm M_{\odot}}$) end of the IMF (for example, changing to a Chabrier 2003 IMF) the effect on the production efficiency is minimal.

The total/integrated spectral energy distribution (SED) of each galaxy is determined by assigning a pure stellar\footnote{i.e. not including the effects of nebular continuum and line emission.} simple stellar population (SSP) SED to every star particle, using the ages and metallicities of the individual particles.

\begin{table*}
\caption{The stellar population synthesis models considered along with the average luminosity-weighted value of the production efficiency $\xi_{\rm ion}/({\rm erg^{-1} Hz})$ in galaxies with $M_*>10^{8}\,{\rm M_{\odot}}$ and the dependence of the $\xi_{\rm ion}$ on redshift and stellar mass. \label{tab:main_tab}}
\begin{tabular}{lclccc}
\hline
Model & vs. & Reference(s) & $<\log_{10}\xi>$ & ${\rm d}\log_{10}\xi/{\rm d}z$ & ${\rm d}\log_{10}\xi/{\rm d}\log_{10}M_*$\\
\hline
{\sc pegase} & 2 & Fioc \& Rocca-Volmerange 1997,1999  & 25.16 & 0.017 & -0.04\\
BC03 & & Bruzual \& Charlot (2003)  & 25.15 & 0.017 & -0.03\\
M05 & & Maraston (2005)  & 25.11 & 0.007 & -0.01\\
{\sc fsps} & 2.4 & Conroy, Gunn, \& White (2009); Conroy \& Gunn (2010)  & 25.25 & 0.020 & -0.06\\
{\sc bpass}/single & 2 & Stanway, Eldridge, \& Becker (2015); Eldridge et al. {\em in-prep} & 25.29 & 0.018 & -0.05\\
{\sc bpass}/binary & 2 & Stanway, Eldridge, \& Becker (2015); Eldridge et al. {\em in-prep} & 25.51 & 0.015 & -0.09\\
\hline
\end{tabular}
\end{table*}

\section{The Lyman Continuum Photon Production Efficiency}\label{sec:xi}

Using the integrated spectral energy distribution we determine the production efficiency $\xi_{\rm ion}$ for each galaxy,
\begin{equation}
\xi_{\rm ion}/({\rm erg^{-1}\, Hz}) = \int_{\infty}^{c/91.2{\rm nm}} L_{\nu}\,(h\nu)^{-1} {\rm d}\nu / L_{\nu}(0.15\mu {\rm m}).
\end{equation}

\begin{figure}
\centering
\includegraphics[width=20pc]{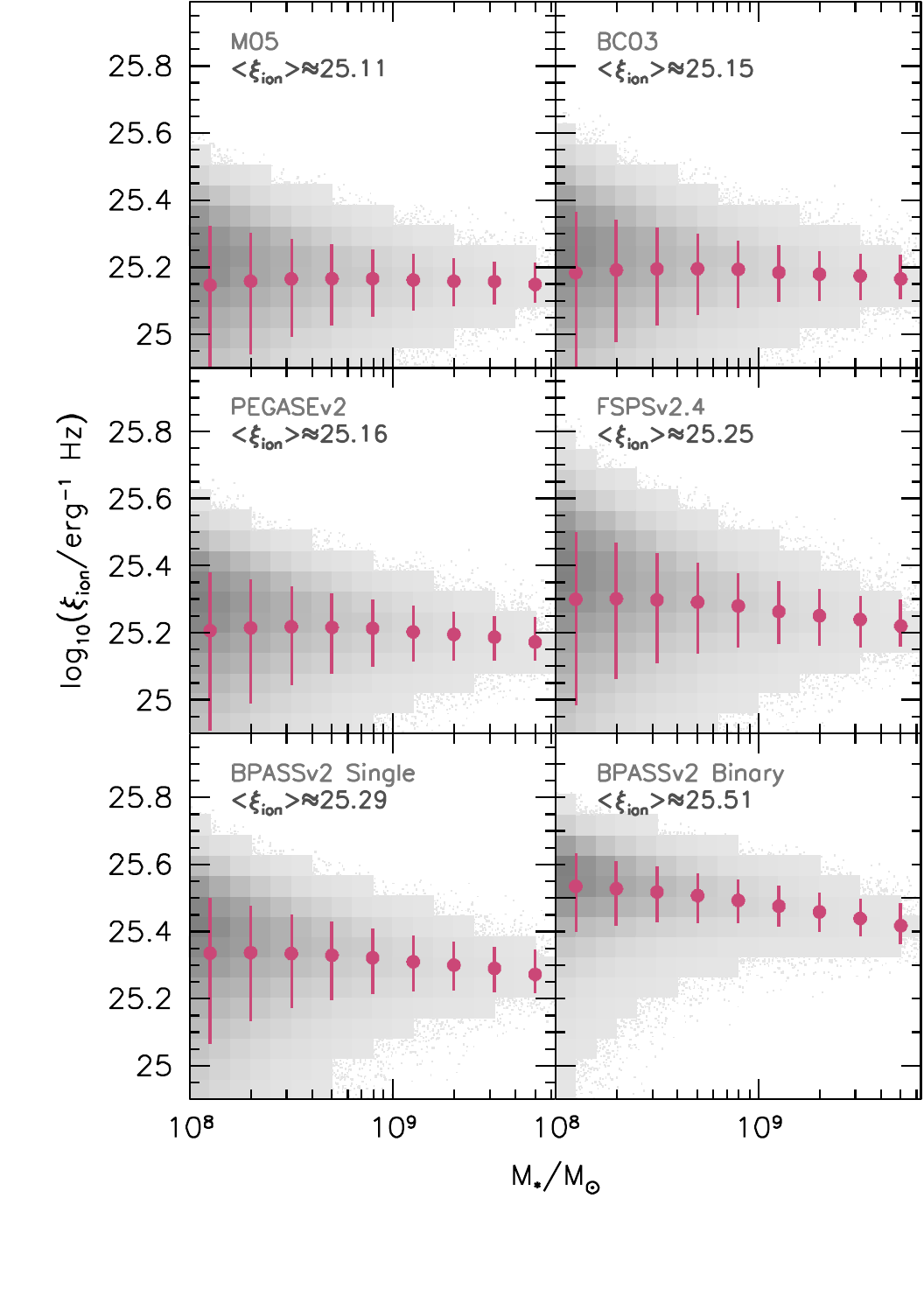}
\caption{The LyC photon production efficiency assuming several different SPS models as function of the stellar mass in the BlueTides simulation at $z=8$. The greyscale blocks show a 2D histogram of the distribution of values of $\xi_{\rm ion}$. Where there are $<10$ sources in a bin, the sources are plotted individually. The points denote the median in each mass bin while the error bars denote the $16-84$ centile range.}
\label{fig:efficiency_BT}
\end{figure}

The distribution of the production efficiences and stellar masses of galaxies with $M_*>10^8\,{\rm M_{\odot}}$ in the BlueTides simulation at $z=8$ are shown in Figure \ref{fig:efficiency_BT}. Each panel shows the assumption of a different SPS model. We also calculate the average luminosity-weighted value of $\xi_{\rm ion}$ for galaxies with $M_*>10^{8}\,{\rm M_{\odot}}$ for each model and present these in Table \ref{tab:main_tab}.

\subsection{Sensitivity to Choice of SPS model}\label{sec:xi.SPS}

\begin{figure}
\centering
\includegraphics[width=20pc]{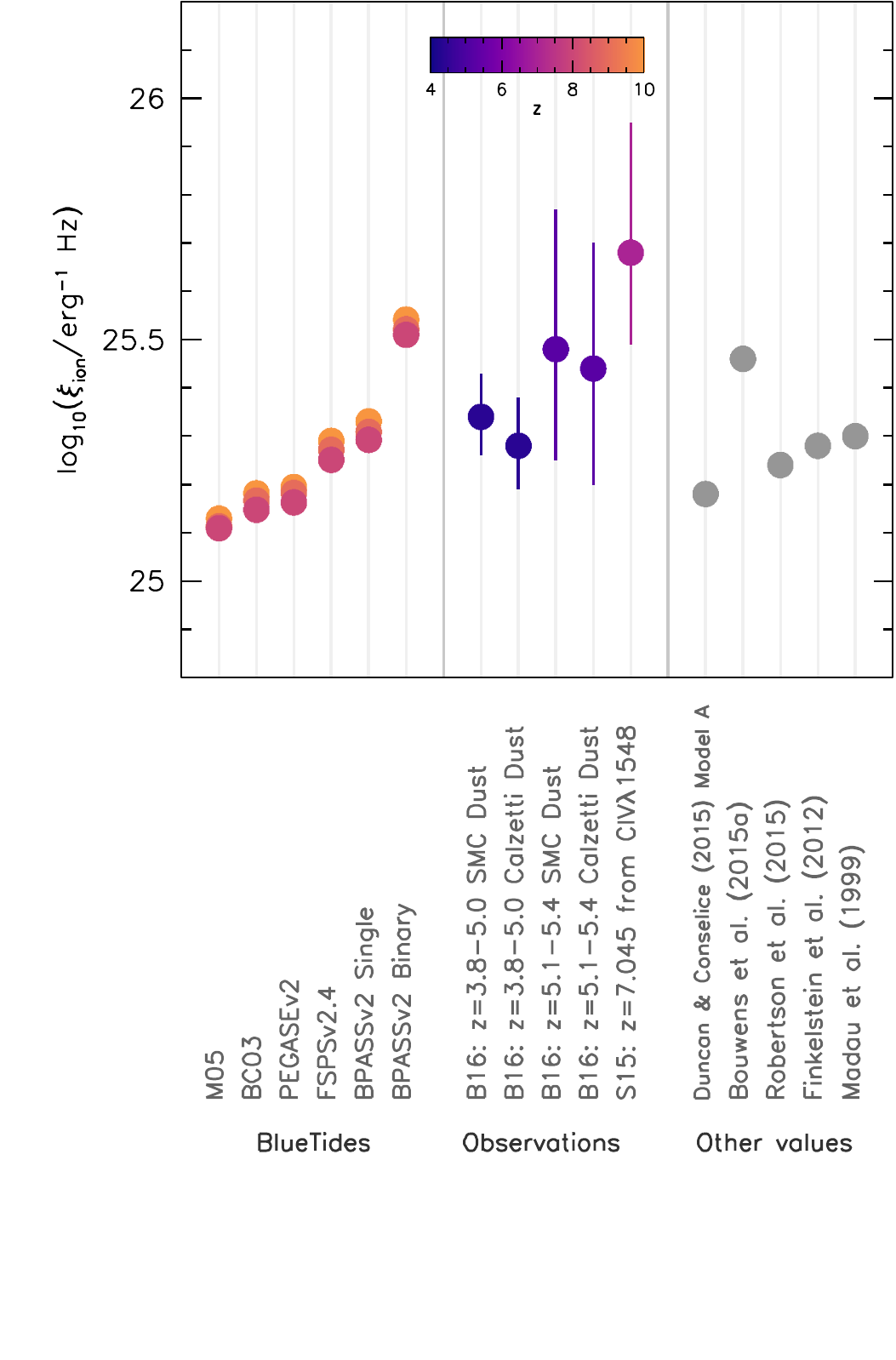}
\caption{Summary of theoretical and observational constraints on the production efficiency $\xi_{\rm ion}/({\rm erg^{-1}\, Hz})$. The grey points denote values assumed in the literature. Observational constraints from Bouwens et al. (2016) and Stark et al. (2015) are based on an assumed LyC escape fraction of zero. An escape fraction of $f_{\rm esc, LyC}=0.1-0.2$ will increase the observationally inferred value of the production efficiency by $0.02-0.07\,{\rm dex}$.}
\label{fig:summary}
\end{figure}

Evident in Figure \ref{fig:efficiency_BT} is that the largest effect on the prediction production efficiency is the choice of SPS model. This can be seen more clearly in Figure \ref{fig:summary} where we plot the average luminosity weighted value of $\xi_{\rm ion}$ (for galaxies with $\log_{10}(M_*/{\rm M_{\odot}})>8$) for each model. The values of $\xi_{\rm ion}$ range from $\log_{10}(\xi_{\rm ion}/{\rm erg^{-1}\, Hz})\approx 25.1-25.5$ with the smallest average value ($\approx 25.11$) inferred assuming the Maraston (2005) model and the largest value from the {\sc bpass} binary model ($\approx 25.51$). Put another way, this implies we would infer $\times 2.5$ as many ionising photons from the same observed UV luminosity assuming the {\sc bpass} binary model compared to the Maraston (2005) model. 

As noted previously, these differences reflect the choice of evolution and atmosphere models in each code. The significantly higher production efficiency obtained using the binary scenario of {\sc bpass} model reflects the impact of binary interactions. These effects are discussed in more detail in Stanway, Eldridge, \& Becker (2015). 

It is important to note that the models considered in this work do not fully encompass the range of potential evolution and atmosphere models. For example, Topping \& Shull (2015) consider the effect of rotation (in addition to the metallicity and IMF) on the production efficiency. They find that by including rotation the production efficiency can be increased.

\subsection{Sensitivity to stellar mass}\label{sec:xi.smass}

It is useful to consider whether there is significant trend of $\xi_{\rm ion}$ with stellar mass. We quantify this using linear regression to calculate ${\rm d}\log_{10}(\xi_{\rm ion}/{\rm erg^{-1}\, Hz})/{\rm d}\log_{10}(M_*/M_{\odot})$ for each SPS model and present these values in Table \ref{tab:main_tab}. For each choice of SPS model there is systematic trend for lower values of $\xi$ at higher masses albeit in most cases very weak. The weakest trend is found using the M05 model (${\rm d}\log_{10}(\xi_{\rm ion}/{\rm erg^{-1}\, Hz})/{\rm d}\log_{10}(M_*/M_{\odot})\approx -0.01$) while the strongest trend comes from the {\sc bpass} binary model (${\rm d}\log_{10}\xi_{\rm ion}/{\rm d}\log_{10}(M_*/M_{\odot})\approx -0.09$). This reflects the stronger sensitivity of $\xi_{\rm ion}$ to the metallicity in these models combined with a strong relationship between the stellar mass and metallicity in the simulation. 

Also notable is an increase in the scatter at low stellar masses; this is consistent as being driven predominantly by the poor sampling of the recent star formation history in these systems.

\subsection{Redshift Evolution}\label{sec:xi.z}

The trend to lower average ages and metallicities at higher-redshift results in higher values of the production efficiency. This can be seen in Figure \ref{fig:summary} where we show the average value of $\xi_{\rm ion}$ at $z\in\{8,9,10\}$ for each SPS model. This evolution is, however, relatively weak with ${\rm d}\log_{10}(\xi_{\rm ion}/{\rm erg^{-1}\, Hz})\xi/{\rm d}z\approx 0.01-0.02$. Values for each SPS model are presented in Table \ref{tab:main_tab}.

\subsection{Comparison to Observations}

There are now a small number of observational constraints on the production efficiency available at high-redshift (albeit at redshifts less than those simulated by BlueTides). Figure \ref{fig:summary} summarises the predictions from BlueTides alongside the observational constraints from Bouwens et al. 2016 and Stark et al. 2015. In deriving these constraints the LyC escape fraction is assumed to be zero. Choosing an LyC escape fraction broadly consistent with the reionisation history of the Universe (i.e. $f_{\rm esc, LyC}=0.05-0.15$) will increase the observationally inferred production efficiency by $0.02-0.07\,{\rm dex}$. While the observational uncertainties are very large, the tend to favour higher values of $\xi_{\rm ion}$ than predicted assuming the M05, BC03, and {\sc pegase} models, despite lying at lower-redshift.


\section{Conclusions}\label{sec:c}

In this work we have coupled a large ($(400/h)^3\,{\rm cMpc^{3}}$) cosmological hydrodynamic simulation (BlueTides) with six different stellar population synthesis models to predict the Lyman-continuum photon production efficiency $\xi_{\rm ion}$. At $z=8$ we find $\log_{10}(\xi_{\rm ion}/{\rm erg^{-1}\, Hz})\approx 25.1-25.5$ (a range of $\times 2.5$) depending on the choice of model. These values are broadly consistent with recent observational constraints ($\log_{10}(\xi_{\rm ion}/{\rm erg^{-1}\, Hz})\approx 25.5$) though somewhat favour the models responsible for the largest values of $\xi_{\rm ion}$, specifically the Binary Population and Spectral Synthesis ({\sc bpass}) model.

As the {\em James Webb Space Telescope} and upcoming generation of ground-based Extremely Large Telescopes become available observational constraints on $\xi_{\rm ion}$ will dramatically improve thanks to rest-frame optical spectroscopy of high-redshift galaxies becoming ubiquitous.

\subsection*{Acknowledgements}

We would like to thank J.J. Eldridge for useful conversations and providing additional models extending {\sc bpass} to lower metallicities. We acknowledge funding from NSF ACI-1036211 and NSF AST-1009781. The BlueTides simulation was run on facilities at the National Center for Supercomputing Applications. SMW and ERS acknowledge support from the UK Science and Technology Facilities Council.

\bsp

\end{document}